\documentclass[prd,preprint,superscriptaddress,showpacs,byrevtex]{revtex4}
\usepackage{latexsym}
\usepackage[dvips]{graphicx}
\begin{document}

\title{An O(N) symmetric extension of the Sine-Gordon Equation }

\author{ Fred Cooper} \email{fcooper@nsf.gov}
\affiliation{T-8, Theoretical Division, MS B285, Los Alamos National
Laboratory,
Los Alamos NM 87545}
\affiliation{National Science Foundation, Arlington, VA 22230}
\author{Pasquale Sodano}  \email{pasquale.sodano@pg.infn.it}
\affiliation{ Dipartimento di Fisica e Sezione I.N.F.N., Universita
di Perugia ,Via A. Pascoli I-06123, Perugia, Italy.}
\author{Andrea Trombettoni} \email{Andrea.Trombettoni@pg.infn.it}
\affiliation{ Dipartimento di Fisica e Sezione I.N.F.N., Universita
di Perugia, Via A. Pascoli I-06123, Perugia, Italy.}
\author{Alan Chodos} \email{chodos@aps.org}
\affiliation{American Physical Society, One Physics Ellipse, College Park, MD
20740} \date{\today}


\begin{abstract}
We discuss an $O(N)$ exension of the Sine-Gordon (S-G)equation described by
the
Lagrangian $ {\cal L} = \frac{1}{2} (\partial_\mu \vec{\phi}) ^2$ + N
$\frac{\alpha_0} {\beta^2} \cos \beta \sqrt {\rho}  $,  where $\rho = \frac
{\vec{\phi} \cdot \vec{\phi}}{N} $ which allows us to  perform an expansion
around the leading order in large-N result using Path-Integral methods.  In
leading order we show our methods agree with the results of a variational
calculation at large-N. We discuss the striking differences for a
non-polynomial interaction between the form for $\langle V[\phi] \rangle$ in
the Gaussian approximation that one obtains at  large-N when compared to the
$N=1$ case.  This is in contrast to  the case when $V[\phi]$ is a
polynomial and no such drastic differences occur. We find for our large-N
extension of the Sine-Gordon model that  the unbroken ground state is
unstable
as one increases the coupling constant (as it is for the original S-G
equation) and we find in leading order that the unbroken symmetry vacuum is
stable as long as $ \beta^2 \leq 24 \pi$.

\end{abstract}

\pacs{ 25.75.Ld, 05.70.Ln, 11.80.-m, 25.75.-q  \hfill
LA-UR-03-1823}

\maketitle
\section{Introduction}
Although the Sine-Gordon equation has previously been studied using
variational
methods, \cite{ref:Coleman} \cite{ref:Boy}
a systematic expansion around the mean field theory result has been lacking.
Here we use the method of auxiliary fields to study going beyond mean field
theory using a systematic expansion based on the the parameter $N$ obtained
when we consider a particular O(N) extension of the orginal S-G model.
Using the methodology of \cite{div}   \cite{ref:Eyal} we first extend the
problem to having O(N) symmetry and then introduce an auxiliary field $\rho=
\frac {\phi_i \phi_i}{N}$ by the insertion of a functional delta function
into the path integral. The large-N expansion is then obtained by
integrating
out exactly the $\phi$ field and then performing a steepest descent
evaluation
of the remaining path integral.  Unlike the case of polynomial interactions,
at leading order in the Gaussian approximation there is a distinct
difference
in the equations of motion when $N=1$ and $N$ large.  We display this
by comparing the $\phi^6$ results with those of the S-G equation.
As in the original S-G equation, our $O(N)$ extended model has an unstable
vacuum as a function of the coupling constant.  We find that in leading
order
in large-N that  the vacuum is unstable for $ \beta^2 \geq 24 \pi$.

\section{Time Dependent Variational Approach}
  One method for obtaining a time dependent
variational approximation to a quantum field theory is to
start with Dirac's Variational principle \cite{ref:Dirac} \cite{ref:Hartree}for
obtaining the functional Schrodinger Equation. By using wave functionals which
become ``exact'' in the
large-N limit-- namely Gaussians we will obtain the time-dependent Hartree
approximation to the exact field equations.

 Dirac's
variational principle
\begin{equation}
\Gamma = \int dt < \Psi | i {\partial
\over \partial t} -H| \Psi >
\end{equation}
leads directly to the Schrodinger equation:
\begin{equation}
 \delta\Gamma = 0 \rightarrow
\{ i {\partial \over \partial t} - H \} |\Psi > = 0
\end{equation}
In the $\varphi$ representation
\[ \Psi[\varphi,t] = < \varphi |\Psi > \]
and
\begin{equation}
H= \int d^n x [ -{1 \over 2} \delta^2/[\delta \varphi(x)]^2+ {1 \over 2}
\nabla
\varphi(x) \nabla \varphi(x) + V[\varphi] ]
\end{equation}

In this representation one can choose a Gaussian
trial wave functional:
\begin{eqnarray}
&&<\varphi|\Psi_{v}> =  \psi_{v}[\varphi,t]=  A \exp \{ -\int_{x,y} \{[\varphi
(x)-  \phi_c(x,t)] \nonumber \\
 &&[G^{-1}(x,y,t)/ 4-i\Sigma(x,y,t)][\varphi(y)-
 \phi_c (y,t)] \} + i  \int_{x}  \pi_c(x,t) [\varphi(x)-  \phi_c(x,t)] \}
\label{eq:trial},
\end{eqnarray}
where
$ \phi_c(x,t)= \langle\Psi_{v}|\varphi (x) | \Psi_{v} \rangle $;
$\pi_c(x,t) =
\langle \Psi_{v}|-i \delta/\delta\varphi(x)| \Psi_{v} \rangle$
and
\begin{equation}
G(x,y,t) = \langle \Psi_{v}|
\varphi(x)\varphi(y)| \Psi_{v} \rangle -  \phi_c(x,t)  \phi_c(y,t) .
\end{equation}

 Then the effective action for the trial wave functional is
\begin{eqnarray}
\Gamma( \phi_c,  \pi_c,G,\Sigma) &&=\int dt < \Psi_{v}| i
\partial/ \partial t -H| \Psi_{v}> \nonumber \\
&&= \int dt dx[\pi_c(x,t)\partial \phi_c(x,t)/ \partial t +\int dt dx dy
\Sigma(x,y;t )\partial G(x,y; t)/\partial t ]\nonumber \\
&&-\int dt < H >
\end{eqnarray}
where
\begin{eqnarray}
< H >&& = \int dx \{  {1 \over 2} \pi^{2} + 2 [\Sigma G\Sigma](x,x) +{1
\over 8}
G^{-1}(x,x) + {1 \over 2} (\nabla \varphi)^{2} \nonumber \\
&& + {1 \over 2}
\lim_{x \rightarrow y}  \nabla_x \nabla_y G(x,y)  + <V> \}.\label{eq:action}
\end{eqnarray}

$< H >$ is a constant of the motion and is a first integral of the motion.
The equations one gets by varying the effective action with respect to
the variational parameters are:
\begin{eqnarray}
\dot{\pi}_c(x,t)&&= \nabla^{2}\phi_c(x,t)- {\partial <V > \over \partial
\phi_c };\nonumber \\ \dot{\phi}_c (x,t) &&= \pi_c(x,t)  \nonumber \\
\dot{G}(x,y; t)&&= 2\int dz[\Sigma (x,z)G(z,y)+G(x,z)\Sigma (z,y)] \nonumber
\\
\dot{\Sigma}(x,y;t)&&=  \int dz[-2 \Sigma (x,z)\Sigma (z,y) + {1 \over 8}
G^{-1}
(x,z)G^{-1} (z,y) ]\nonumber \\
&& +  [{1 \over 2}\nabla^{2}_{x}
- {\partial < V > \over \partial G}] \delta(x-y)  \nonumber \label{eq:lgN}
\end{eqnarray}
For calculating expectation values,
\begin{equation}
P[\phi] = A^2 \exp \{ -\int_{x,y}[\varphi
(x)-  \phi_c(x,t)]
[\frac {G^{-1}(x,y;t)}{2}
[\varphi(y)-
 \phi_c (y,t)]]
\end{equation}
with the normalization $A$ determined from the conservation of probability:
\begin{equation}
\int  {\cal D} \phi~~ P[\phi] = 1
\end{equation}

To get a general expression for $ \langle V \rangle$  one can use the
method of the generating functional:

\begin{equation}
Z[j] = \langle \exp[\int j(x,t) \{ \phi(x,t) - \phi_c(x,t) \} ] \rangle = \exp
[\frac{1}{2} \int j(x,t) G(xy,t) j(y,t)].
\end{equation}
In the single field case (N=1) for example one has
\begin{equation}
\langle (\phi(x,t)-\phi_c(x,t)) ^{2n} \rangle = \frac{\delta ^{2n}
Z[j]}{\delta j^{2n} (x)} |_{j=0}.
\end{equation}
and we obtain the simple result
\begin{equation}
\langle (\phi(x,t)- \phi_c(x,t)) ^{2n} \rangle = (2n-1)!!  G(x,x;t)^{n}
\end{equation}
To obtain the expectation value of an arbitrary potential we assume that the
potential
has a Taylor series expansion about  the classical  field $\phi_c = \langle
\phi \rangle$ and  the expectation value is taken with respect to the
Gaussian
trial wave function.   At N=1 we obtain the result:
\begin{equation}
\langle V[\phi] \rangle = \sum_n \frac{1}{(2n)!} \frac {d^{2n} V[\phi_c]}{d
\phi_c^{2n}} \langle (\phi-\phi_c)^{2n} \rangle = \sum_n \frac{1}{2^n n!}
G^n  \frac {d^{2n} V[\phi_c]}{d \phi_c^{2n}}
\end{equation}
For the O(N) extension of the potential, $ \phi \rightarrow \phi_i;~~~i=1
\ldots N $ and the Green's function $G$ is now an  $N \times  N$ matrix. One
then finds making the Gaussian ansatz and ignoring non-leading terms in
$1/N$ that (see \cite{ref:BMB})
 \begin{equation}
\langle ( \frac {\vec \phi \cdot \vec \phi}{N})^n \rangle \rightarrow
\langle
\rho \rangle^n \end{equation}
where
\begin{equation}
\langle \rho \rangle =  \frac {\langle \vec \phi \cdot  \vec \phi
\rangle}{N}
= G + \frac {\vec \phi_c \cdot \vec \phi_c}{N} \equiv G+\phi_c^2  .
\end{equation}

Thus at leading order in large-$N$ the expectation value of the potential
becomes  \begin{equation}
\langle V[\frac {\vec \phi \cdot \vec \phi}{N} ] \rangle = V_{cl} [ \phi_c^2
+
G ] \end{equation}
\subsection {$\phi^6$ Model}
First let us look at a typical polynomial model such as
the $\phi^6$ model described by the classical potential
\begin{equation}
V_{cl}[\phi] = \frac{\mu^2}{2} \vec \phi \cdot \vec \phi + \frac
{\lambda_0}{4
N} (\vec \phi \cdot \vec \phi)^2 + \frac{\eta}{6 N^2} (\vec \phi \cdot \vec
\phi)^3
 \end{equation}
For $N=1$ one gets the Hartree approximation result
\begin{eqnarray}
\langle V \rangle && = \frac{\mu^2}{2} [\phi^2 + G] + \frac {\lambda_0}{4}
[ \phi^4 + 6 \phi^2 G + 3 G^2 ] \nonumber \\
&& +  \frac{\eta}{6} [\phi^6+ 15 \phi^4 G + 45 \phi^2 G^2 + 15 G^2]
\end{eqnarray}
where in the above $G= G(x,x)$ and $G(x,y;t)$ is the variational function.

To obtain the phase structure  one calculates the
effective potential which is the expectation value of the Hamiltonian in the
Gaussian trial state for spatially homoogenous time independent fields:
\begin{equation}
V_{eff}= \frac{G^{-1}(x,x)}{8}+ \frac{1}{2} lim_{x \rightarrow y} \nabla_x
\nabla_y G(x,y) + \langle V \rangle.
\end{equation}

We can change variables from $G(x,y)$ to the effective mass $\chi$ for
constant (homogeneous) external sources and in that case parametrize
$G(x,y)$
via
\begin{equation}
G(x,y) = \int \frac{dk} {2 \pi} ~~e^{i k(x-y)} ~~\frac {1} {2
\sqrt{k^2+\chi}},
\end{equation}
So that if we introduce a cutoff $\Lambda$
\begin{equation}
G(x,x) = \frac{1}{4 \pi} \ln \frac{\Lambda^2}{\chi}
\end{equation}
The first two terms in the effective potential coming from the kinetic
energy
can now be written as
\begin{equation}
\langle KE \rangle = \int \frac{dk} {2 \pi} [ \frac {\sqrt{k^2+\chi}}{4} +
\frac{k^2} {4 \sqrt{k^2+\chi}} ] \end{equation}
This has an infinite $\chi$ independent term which is the cosmological
constant
and needs to be subtracted by hand.

In general $V_{eff}$ is a functional of $\phi, G$ or $\phi, \chi$. Using the
chain rule we have
\begin{equation}
\frac{\partial V} {\partial \chi} = \frac{\partial V} {\partial
G}\frac{\partial G} {\partial \chi}
\end{equation}
with
\begin{equation}
\frac{\partial G} {\partial \chi} = - \frac {1} {4 \pi \chi}
\end{equation}
 Here
\begin{eqnarray}
\frac {\partial V} {\partial G}&& = \frac{1}{2}[ -\chi + \mu^2 + 3 \lambda
\{ \phi^2 + G[\chi] \} \nonumber \\
&&+ 5 \eta \{ \phi^4 +6  \phi^2 G [\chi] + 3 G^2[\chi]\} ]  \label{gap1a}
\end{eqnarray}

Another way of writing the Kinetic energy which removes the cosmological
constant
term is used in \cite{ref:BMB} and leads to a result independent of the
cutoff.
There one lets  $\chi \rightarrow m^2$ in the definition of $G$  and sets
  \begin{equation}
K[\chi] = -\frac{1}{2} \int_0^\chi m^2  \frac {\partial G[m^2]}{\partial
m^2}
= \frac {\chi}{8 \pi}.
 \end{equation}

The potential energy also has divergences which are related to mass and
coupling constant renormalization if one does not do any normal ordering.
Here
we will first use conventional renormalization and then show it is
equivalent
(in 1+1 dimensions) to normal ordering the unrenormalized result with
respect
to the renormalized mass.   First let us look at the case $N=1$.
We define the renormalized mass  as the value of $\chi$
determined from the gap equation obtained from setting  $\frac {\partial V}
{\partial G}|_{\phi=0} =0 $  i.e.  \begin{equation}
m^2  = \mu^2 + 3 \lambda_0 G[m^2]+ 15 \eta G^2 [m^2]
\end{equation}
The gap equation for $\chi$ is fully renormalized if we solve the above for
$\mu$ in terms of $m$ and also perform the coupling constant renormalization
\begin{equation}
\lambda_r= \lambda_0 + 10 \eta G[m^2]
\end{equation}
Defining
\begin{equation}
G_r[\chi] = G[\chi]-G[m^2] = \frac{1}{4 \pi} \ln \frac{m^2}{\chi}
\end{equation}
one then obtains the renormalized equation
\begin{eqnarray}
\frac {\partial V} {\partial G}&& = \frac{1}{2}[ -\chi + m^2 + 3 \lambda_r
\{ \phi^2 + G_r[\chi] \} \nonumber \\
&&+ 5 \eta \{ \phi^4 +6  \phi^2 G_r [\chi] + 3 G_r^2[\chi]\} ]  \label{gap2}
\end{eqnarray}

We notice here by comparing Eq. \ref{gap1a} and Eq. \ref{gap2} that the
{\em only} difference between the unrenormalized and renormalized gap
equations
are the replacement of:
\begin{equation}
\mu \rightarrow m; ~~ \lambda \rightarrow \lambda_r;  G \rightarrow G_r.
\end{equation}

This is an example of the result in two space-time dimensions (see also
\cite{ref:chang}) that if one did {\em no} mass and coupling constant
renormalization, but normal ordered the two point function with respect to the
renormalized mass parameter $m$ the theory would be rendered finite.

To obtain the renormalized effective potential one needs to do an
integration of the renormalized equaition for $\frac{\partial V}{\partial
\chi}$ and add  the pure $\phi$ dependent terms.

We have
\begin{eqnarray}
\frac {\partial V}{\partial \chi}&& = - \frac{1}{8 \pi \chi} [ -\chi + m^2 +
3 \lambda_r
\{ \phi^2 + G_r[\chi] \} \nonumber \\
&&+ 5 \eta \{ \phi^4 +6  \phi^2 G_r [\chi] + 3 G_r^2[\chi]\} ]  \label{gap3}
\end{eqnarray}
Integrating and adding the pure $\phi$ dependent terms we obtain:
 \begin{eqnarray}
V[\phi,\chi] && = \frac{1}{2} m^2 \phi^2 + \frac{1}{4} \lambda_r \phi^4
+\frac{1}{6} \eta \phi^6
\nonumber \\
&& \frac{1}{8 \pi} [ \chi - m^2 + (m^2+ 3 \lambda_r \phi^2+ 5 \eta \phi^4)
\ln \frac{m^2}{\chi} ]
\nonumber
\\
&&\frac{3}{64 \pi^2} (\lambda_r  + 10 \eta \phi^2) \ln^2 \frac{m^2}{\chi}
+\frac{5}{128 \pi^3} \eta
\ln^3
\frac{m^2}{\chi}.
\end{eqnarray}

\subsection{ $\phi^6 $at large $N$}
At leading order in large $N$ one has as discussed before that
\begin{equation}
\langle V \rangle /N = \frac{1}{2} \mu^2  [\phi_c^2+G(x,x)]
+\frac{\lambda_0}{4}  [\phi_c^2+G(x,x)]^2+
\frac{\eta}{6}
 [\phi_c^2+G(x,x)]^3,
\end{equation}
 where here we use the shorthand $ \phi_c^2 \equiv \frac {\vec \phi \cdot
\vec
\phi} {N}$.  Performing the mass renormalization we find now:

\begin{equation}
m^2 = \mu^2 + \lambda_0 G(x,x;m^2) + \eta_0 G^2(x,x;m^2).
\end{equation}
The equation for the coupling constant renormalization is also slightly
changed and reads:
\begin{equation}
\lambda_r = \lambda_0 + 2 \eta G(x,x;m^2)
\end{equation}
then the renormalized gap equation is
\begin{equation}
\frac {\partial V}{\partial G} = \frac{1}{2} [ -\chi + m^2  + \lambda_r \{
\phi^2 + G_r(x,x;\chi) \} + \eta  \{
\phi^2 + G_r(x,x;\chi)\}^2] .
\end{equation}
Thus apart from the factors in front of $\lambda$ and $\eta$  reflecting $3=
1+2/N$ etc, the renormalization at large $N$ is quite similar to what
happens
in the Hartree approximation.
We also find
 \begin{equation}  \frac{\partial
H} {\partial \chi} = \frac{\partial H} {\partial G} \frac{\partial G}
{\partial
\chi}  \end{equation} which after renormalization can be written (using
$V_{eff}$ for $H$ at constant fields)
\begin{equation}
\frac{\partial V_{eff}} {\partial \chi}= \frac{1} {8 \pi \chi} \left[\chi -
m^2 - \lambda_r \left(\phi^2
+\frac{1}{4
\pi} {\ln} [\frac{m^2}{\chi} \right) - \eta \left( \phi^2 +\frac{1}{4 \pi}
{\ln} [\frac{m^2}{\chi}] \right)^2  \right]
\end{equation}
This result  again could have been obtained if we did {\em NOT}
renormalize the coupling constant or the  mass and instead used $m^2$ for
the
mass parameter and everywhere made the subtraction: \begin{equation}
G(xx,\chi) \rightarrow G(xx, \chi) - G(xx, m^2) \equiv G_r(x,x;\chi).
\end{equation}
Integrating as before we find the renormalized effective potential can
be written in the form
\begin{eqnarray}
V[\phi,\chi]/N =&&\frac{1}{8 \pi} [ \chi - m^2]  + \frac{1}{2} m^2 [\phi^2
+G_r[\chi]] \nonumber \\
&& + \frac{1}{4} \lambda_r [\phi^2+G_r]^2 + \frac{1}{6} \eta[\phi^2 + G_r]^3
\end{eqnarray}
which has the advertised form.
\subsection{sine-Gordon equation}

The sine-Gordon equation is quite interesting since the classical
equation has a kink solution which remains after quantization and  there is
also
a phase transition as we increase the effective coupling constant as
discussed
by Coleman \cite{ref:Coleman}.  The  usual sine-Gordon equation is described
by
the Hamiltonian: \begin{equation}
H= \int dx [ {1 \over 2} \pi(x,t)^2 + {1\over 2} (\nabla \varphi(x,t))^2
- {\alpha_0 \over \beta^2} \cos \beta \varphi + \gamma ]
\end{equation}
where in the ``$\varphi$" representation
\[
\pi(x) = -i {\delta \over \delta \varphi}
\]

Taking the expectation value of this Hamiltonian in the trial
Gaussian wave functional state one now has
\begin{equation}
< V > = < - {\alpha_0 \over \beta^2} \cos \beta \varphi > =
  - {\alpha_0 \over \beta^2} \cos \beta \phi~~ e^{- {\beta^2 \over 2}
G(x,x)}
\end{equation}

Varying the action we obtain the following field equations for the
variational functions:
\begin{equation}
\ddot{\phi} - \nabla^2 \phi + {\alpha_0 \over \beta} \sin \beta \phi
~~e^{- {\beta^2 \over 2} G(x,x)} =0 \label{eq:phi},
\end{equation}
\begin{equation}
\dot{G}(x,y) =  2 \int dz  \{G(x,z) \Sigma(z,y) + \Sigma (x,z) G(z,y) \}
\label{eq:g}
\end{equation}
\begin{eqnarray}
\dot{\Sigma}(x,y;t)&&=  \int dz[-2 \Sigma (x,z)\Sigma (z,y) + {1 \over 8}
G^{-1}
(x,z)G^{-1} (z,y) ]  \nonumber \\
&& +  [{1 \over 2}\nabla^{2}_{x}
-{1\over 2} \alpha_0 \cos \beta \phi(x,t)
~~e^{- {\beta^2 \over 2} G(x,x)}] \delta (x-y)
\label{eq:sig2}
\end{eqnarray}

In the Heisenberg picture, making the same the Hartree approximation gives
the
following  covariant equation for the nonequaltime correlation function:

\begin{equation}
[\Box + \langle \frac{\partial^2 V}{\partial \phi^2} \rangle ]{\cal G}
(x,x') = \delta^2(x-x')
\end{equation}

where
\begin{equation}
\langle \frac{\partial^2 V}{\partial \phi^2} \rangle = \alpha_0 \cos \beta
\phi(x,t)
~~e^{- {\beta^2 \over 2} G(x,x)} \equiv \chi(x).
\end{equation}
Again the connection between the covariant and noncovariant Green's
functions
at equal spatio-temporal points is: $ \frac {{\cal G}(x,x; t,t)}{i} =
G(x,x;t)$.

  When $\chi$ is
independent of $x$ one can again introduce the Fourier Transform:
\begin{equation}
G(x,y) = {1 \over 2 \pi} \int dk{\tilde G}(k) e^{i k(x-y)}
\end{equation}
where now
\[ \tilde{G} (k) = {1 \over 2 (k^2 + \chi)^{1/2}} \]
In particular, when $\phi=0$ one has the gap equation:
\begin{equation}
\chi = \alpha_0
~~e^{- {\beta^2 \over 2} G(x,x;\chi)} = \alpha_0
(\frac{\Lambda^2}{\chi})^{-\frac{\beta^2}{8\pi}}
\end{equation}

This equation tells one how to choose $\alpha_0$ as a function of the cutoff
$\Lambda$ to
insure  that the physical mass $\chi$ is finite.

To study the phase transition in this theory in this approximation we follow
\cite{ref:Coleman} \cite{ref:Boy}.   We determine for what values of
the coupling constant one can have $\phi=0$ be a minimum of  the energy.

In the vacuum sector we have  for the Hamiltonian Density:
\begin{equation}
< {\cal{H}} > = {1 \over 8} G^{-1}(xx;\chi) + {1 \over 2}
\lim_{x \rightarrow y}  \nabla_x \nabla_y G(x,y;\chi)
 - {\alpha_0 \over \beta^2}~~ e^{- {\beta^2 \over 2} G(x,x;\chi)}
\end{equation}
Inserting a momentum space cutoff $\Lambda$
\begin{equation}
{\cal{H}}(\mu^2) = {1 \over 8 \pi} \int_{-\Lambda}^{\Lambda} dk
{2 k^2 + \mu^2 \over (k^2 + \mu^2)^{1/2} } - {\alpha_0 \over \beta^2}
 ( {\Lambda^2
\over \mu^2})^ {-\beta^2/ 8 \pi}
\end{equation}
Let us define an arbitrary  renormalized mass squared parameter $\alpha_r$
by
\begin{equation}
\alpha_r(m^2) = \alpha_0 ({m^2 \over \Lambda^2})^{\beta^2/ 8 \pi}
\end{equation}
which has the property that
\begin{equation}
\alpha_r(\chi) = \chi
\end{equation}
we find that the once subtracted (at $m^2$) energy density is given by
\begin{equation}
{\cal{H}}(\chi)-{\cal{H}}(m^2)= {1 \over 8 \pi} (\chi - m^2)
- {\alpha_r \over \beta^2} ({\chi \over m^2})^{\beta^2 / 8 \pi}
\end{equation}
The first derivative is given by:
\begin{equation}
\frac{\partial H}{\partial \chi} = \frac{1}{8 \pi} [1-
\frac{\alpha_r(m^2)}{m^2}
(\frac{\chi}{m^2})^{\beta^2/8 \pi -1}]
\end{equation}
The minimum for $\phi=0$ is at
\begin{equation}
(\frac{\chi}{m^2})^{\beta^2/8 \pi -1}= \frac{m^2}{\alpha_r(m^2)}
\end{equation}
We notice that if we choose
\[   m^2 = \alpha_r (m^2)\]
then the extrema of the energy is at
\[  \chi = m^2. \]
Since the second derivative of the energy is given by:
\begin{equation}
{\partial^2 {\cal{H}} \over \partial^2\chi} = -{1 \over 8 \pi m^2}
({\beta^2 \over 8 \pi}-1) ({\chi \over m^2})^{(\beta^2/ 8 \pi)-2}
\end{equation}
we find that a stable ground state with unbroken symmetry ($\phi=0$) exists
only for
\begin{equation}
{\beta^2 \over 8 \pi} < 1
\end{equation}
which is in accord with the result of Coleman \cite{ref:Coleman}.

The resulting renormalized effective potential is
\begin{equation}
V[\phi,\chi] = \frac{1}{8 \pi} [\chi-m^2] - \frac{m^2}{\beta^2} (\frac
{\chi}{m^2})^{(\beta^2/ 8 \pi)}
\cos \beta \phi
\end{equation}

For the Schrodinger picture update equations we can defined a finite
spatially and temporally varying renormalized mass $M^2(x,t)$ via
\begin{equation}
M^2(x,t) \equiv   m^2 \exp[-{\beta^2 \over
2}\{G(xx;\chi)-G(xx;m^2) \} ]= m^2 \exp[-{\beta^2 \over
2}G_r(xx;\chi)]
\end{equation}

This mass renormalization will render the resulting update equations
for the time evolution problem finite.
In terms of the renormalized mass $M^2(x,t)$ the renormalized TDHF equations
are
:

\begin{equation}
\ddot{\phi} - \nabla^2 \phi + {M^2(x,t) \over \beta} \sin \beta \phi
 =0 \label{eq:phir},
\end{equation}
\begin{equation}
\dot{G}(x,y) =  2 \int dz  \{G(x,z) \Sigma(z,y) + \Sigma (x,z) G(z,y) \}
\label{eq:gr}
\end{equation}
\begin{eqnarray}
\dot{\Sigma}(x,y;t)&&=  \int dz[-2 \Sigma (x,z)\Sigma (z,y) + {1 \over 8}
G^{-1}
(x,z)G^{-1} (z,y) ]  \nonumber \\
&& +  [{1 \over 2}\nabla^{2}_{x}
-{1\over 2} M^2(x,t) \cos \beta \phi(x,t)
] \delta^{d}(x-y)
\label{eq:sig2r}
\end{eqnarray}

In the Heisenberg picture, the covariant time dependent equations in terms
of
$M^2(x,t)$ are

\begin{equation}
[\Box + M^2(x,t) \cos \beta \phi(x)] {\cal G}(x,x') = \delta^2 (x-x')
\end{equation} thus the space and time dependent effective mass is
\begin{equation}
\chi(x) = m^2 \cos \beta \phi(x) e^{-\frac{\beta^2 {\cal G}_r(x,x;\chi)}{2i}} = M^2
(x,t) \cos \beta \phi(x) \label{eq:space}\end{equation}
The expectation value of the field obeys the equation
\begin{equation}
\Box \phi(x)  + M^2(x,t) \sin \beta \phi(x)       =0.
\end{equation}
The subscript $r$ in Eq. \ref{eq:space} means that one uses the once
subtracted covariant Green's function. Here we see that the main effect of the Hartree approximation on the
kink evolution equation is to replace $m^2$ by a self consistently determined
$M^2(x,t)$. We notice that the "classical" field evolves quite differently
than the propogator.  As we shall
see next, making the Gaussian approximation but keeping
only leading order terms at large-N dramatically alters the evolution
equation
in a manner that can be interpreted as a change in the behavior of the
periodicity itself, but with the
evolution of $\phi$ and ${\cal G}$ being similar. 
For the N-component field we will consider instead the classical potential:
\begin{equation}
V_{cl} = -N \frac {\alpha_0} {\beta^2} \cos \beta \sqrt{\rho};  ~~~~~\rho
\equiv \frac{\vec \phi \cdot \vec \phi} {N}.
\end{equation}
Using the previous result at large $N$ that $ \langle H \rangle = K[\chi] +
V_{cl}\left[ \phi_c^2 + G[\chi]\right]$,
we have that the effective potential is now:
\begin{equation}
V_{eff}/N  = \frac{\chi}{8 \pi} - \frac {\alpha_0} {\beta^2} \cos \beta
\sqrt{\phi_c^2 + G[\chi]}.
\end{equation}
The renormalized effective potental is thus given by
\begin{equation}
V_{eff}/N  = \frac{\chi}{8 \pi} - \frac {\alpha} {\beta^2} \cos \beta
\sqrt{\phi_c^2 + G_r[\chi;m^2]}.
\end{equation}
with $m^2$ an arbitrary subtraction point.
If we choose the extremum of energy when $\phi=0$ to be at $\chi= m^2$ than
the
renormalized effective potential can be written as:
\begin{equation}
V[\phi,\chi] = \frac{1}{8 \pi} [\chi -m^2] -\frac{m^2}{\beta^2} \cos \left(
\beta \sqrt{\phi^2+
\frac{1}{4 \pi} \ln[\frac{m^2}{\chi}]} \right)
\end{equation}

We notice now that the effective mass (pole in the propagator) is  at
\begin{equation}
\chi=
m^2
\frac {\sin \left( \beta \sqrt{\phi^2+ \frac{1}{4 \pi}
\ln[\frac{m^2}{\chi}]} \right)}{\beta \sqrt{
\phi^2 +\frac{1}{4
\pi}
\ln[\frac{m^2}{\chi}]}}
\end{equation}
which is functionally totally different than the $N=1$ Hartree result that
\begin{equation}
\chi(x) = m^2 \cos \beta \phi(x) e^{-\frac{\beta^2 {\cal G}_r(x,x;\chi)}{2i}} = M^2
(x,t) \cos \beta \phi(x) \end{equation}

 \section{Path Integral Approach}

To obtain the large-N expansion for an arbitrary even polynomial interaction
we follow the work of Eyal et. al.
\cite{ref:Eyal}.  Starting from the usual path integral for the generating
functional
\begin{equation}
Z[j] = \int D \phi~~{\rm Exp} \left[ i \{ \frac{1}{2 }[\partial_\mu
\phi_i]^2 - N
V[\frac{\vec \phi \cdot \vec \phi}{N}] +\vec j \cdot \vec \phi \} \right] \end{equation}
one introduces the composite field $\rho$ using :
\begin{equation}
  1= \int d\chi \delta( \rho- \frac{\vec \phi \cdot \vec \phi}{N}) = \int
d\rho
 d \chi ~\exp^{ \left[ i N \frac{\chi}{2} (\rho - \frac{\vec \phi \cdot \vec
\phi}{N}) \right]} \end{equation} This allows us to rewrite the generating
functional (also adding sources for $\chi$ and $\rho$)  as
\begin{equation}
Z = \int D \phi D\chi D \rho{\rm Exp} \left[ i \{  \frac{1}{2} [
\partial_\mu \phi_i ]^2-
\frac{1}{2} \chi {\vec \phi \cdot \vec \phi}- N V[{\rho}] +
\frac{N}{2} \rho \chi +\vec j \cdot \vec \phi + N S \chi + N J
\rho  \} \right] \end{equation}

The equations resulting from the second form of the action are
\begin{equation}
[\Box + \chi] \phi_i = j_i~; ~~~\chi= -2J - 2 V^\prime[\rho]~; ~~~\rho =
\vec \phi \cdot \vec \phi/N - 2 S
\end{equation}

The large-N expansion is obtained by integrating out the $\phi$ fields and,
recognizing that the
resulting action is proportional to $N$
performing the remaining  integral by Stationary Phase.  The first
integration yields
\begin{eqnarray}
&& Z[j,S,J] = \int  D\chi  D  \rho e^{ i N S_{eff}[\rho,\chi, j,S,J]}
\end{eqnarray}
where
\begin{equation}
S_{eff} =  i \{ [  \int  \frac {\vec j(x) \cdot \vec j (y)}{2 N}
{\cal G}[x,y; \chi]
- V[\rho] + \frac{1}{2} \rho \chi+  S \chi + J {\rho} + \frac{i}{2} Tr {\rm
Ln}
{\cal G}^{-1} [x,y, \chi] \} \biggr]
\end{equation}
and here
\begin{equation}
{\cal G}^{-1} [x,y, \chi]= [\Box + \chi] \delta^2(x-y)
\end{equation}
is now the covariant two time Green function.
The stationary phase conditions are:
\begin{eqnarray}
\frac{1}{i}\frac{\delta S}{\delta \chi}&&= S+\frac{\rho}{2} + \frac{i}{2}
{\rm
Tr} {\cal G} - \frac {1}{2}\phi_c^2 =0, \nonumber \\
\frac{1}{i}\frac{\delta S}{\delta \rho}&&= \frac {\chi}{2} - V'[\rho] + J =0
\end{eqnarray}
where $ {\vec \phi}_c  =  \int {\cal G} {\vec j}$ and $\phi_c^2= {\vec \phi}
\cdot {\vec \phi}/N$.
In the absence of sources, this leads to the constraint equation for $\rho$
\begin{equation}
\rho = \phi_c^2 + \frac{1}{i} {\rm Tr }{\cal G}
\end{equation}
and the equation for the effective mass (gap equation) in leading order:
\begin{equation}
\chi = - 2 V'[\rho].
\end{equation}
At leading order we also have from the definition of $\phi_c= \frac{1}{1}
\frac {\delta ln Z}{\delta j}$ that
\begin{equation}
[\Box + \chi] \vec \phi_c = \vec j
\end{equation}
so that unlike the Hartree case the mass for the classical field $\phi$ is
the same as the mass entering into the propagator.
The Gaussian fluctuations about this mean field (Stationary point of the
effective action)  is obtained from the second derivative of the effective
action at the stationary point. This matrix is actually the inverse
propagator
of the two component field made up of $\chi$ and $\rho$.  The stability of
the
large-N approximation is related to the eigenvalues of this matrix being
positive. Critical behavior is related to the deteminant of the matrix being
zero.   The second derivatives are given by
\begin{equation}
\frac{1}{i} \frac {\delta^2 S}{\delta \chi(x) \delta \chi (y)} =
-\frac{i}{2}
{\cal G} (x,y) {\cal G}(y,x) + \phi_c(x) {\cal G}(x,y) \phi_c(y) \equiv
\frac
{\Sigma(x,y)}{2}
\end{equation}
\begin{equation}
\frac{1}{i} \frac
{\delta^2 S} {\delta \chi(x) \delta \rho(y)}=\frac{1}{i} \frac {\delta^2 S}
{\delta \chi(x) \delta \rho(y)} =\frac{1}{2} \delta (x-y)
\end{equation}
\begin{equation}
\frac{1}{i} \frac {\delta^2 S}{\delta \rho(x) \delta \rho (y)}= - V"[\rho]
\delta(x-y).
\end{equation}
For constant fields, the Fourier transform of the inverse propagator is
given
by  \begin{equation}
  D^{-1}[p;\chi,\rho,\phi] =
\left( \begin{array}{cc}
 \frac{\Sigma(p)}{2} & \frac{1}{2} \\
\frac{1}{2} & -V"[\rho]
\end{array}
\right)
\end{equation}

   Expanding $S$ about the stationary phase point, keeping up to Gaussian
fluctuations and Legendre transforming one has the effective action to $1/N$
is
\begin{equation} \frac {\Gamma}{N}=  \int dx {\cal L}_{cl}[\phi,\rho,\chi] +
\frac{i}{2} Tr {\ln} {\cal G}^{-1} [x,y, \chi] \}+\frac{i}{2 N} Tr {\ln}
D^{-1} [x,y, \chi,\rho, \phi] \} \end{equation} and the inverse matrix
propagator $D^{-1}$is defined in terms of the second derivatives of the
effective action at the stationary phase point with respect to the fields
$\rho$ and $\chi$.

In 1+1 dimensions, $\Sigma(q)$ is
finite
and (for the case of zero classical field) given by the integral
\begin{equation}
\Sigma(q; \chi) = \frac{1} {i} \int \frac{d^2 p}{(2\pi)^2} \frac{1}{(-p^2 +
\chi)(-(p-q)^2 + \chi)}= \int \frac{d^2 p_E}{(2\pi)^2} \frac{1}{(p_E^2 +
\chi)(p_E-q_E)^2 + \chi)}
\end{equation}
where in the last equality we have made the Wick rotation $p_0 \rightarrow i
p_{0E}$.  The value of $\Sigma(0)$ is important in studying the critical
points of the theory where the determinant of the quadratic fluctuations
around the mean field becomes zero and one has a massless  excitation.  At
that
point the large-N expansion breaks down and a resummation technique is
needed
as discussed in \cite{ref:Eyal}.  Explicitly \begin{equation}
\Sigma(q=0,\chi) = \frac{1}{4 \pi \chi}.
\end{equation}
and the condition for the determinant to vanish becomes
\begin{equation}
\Sigma(0) V^{\prime \prime} [\rho] + \frac{1}{2} = 0. \label{eq:detzero}
\end{equation}

\subsection{$ \phi^6$ model at large-N}
 We start with the
usual Lagrangian for the $\phi^6$ model is  (see \cite{ref:BMB})
\begin{equation} {\cal L} = \frac{1}{2} (\partial_\mu \phi_i)^2 -
\frac{1}{2}
\mu^2  \vec \phi \cdot \vec \phi  -\frac{\lambda_0}{4N}  (\vec \phi
\cdot \vec \phi )^2-
\frac{\eta}{6 N^2}
 (\vec \phi
\cdot \vec \phi) ^3
\end{equation}
and $ i= 1 \ldots N$.

Introducing the functional $\delta$ function that $\rho= \frac{\vec \phi \cdot \vec
\phi}{N}$ we get the second form of the Lagrangian
\begin{equation} 
\frac {\cal L}{N}  = \frac{1}{2 N} (\partial_\mu \phi_i)^2 -
\frac{1}{2}
\mu^2  \rho  -\frac{\lambda_0}{4} \rho^2-
\frac{\eta}{6 }
\rho^3 + \frac{\chi}{2} (\rho - \frac{\vec \phi \cdot \vec \phi}{N}).
\end{equation}

The effective potential obtained at leading order in large N for $\phi^6$
theory written in terms of
the auxiliary fields $\chi,\rho$ is simply (scaling out the $N$)
\begin{equation}
V_{eff}=  \frac{1}{2} \chi( \phi^2 - \rho)  +\frac{1}{2} \mu^2 \rho + \frac
{\lambda_0}{ 4} \rho^2 +
\frac{\eta}{6} \rho^3 + \frac{1}{2iN} {\rm tr} {\ln} {\mathcal
G}^{-1}[\chi]
\end{equation}
where here ${\mathcal G}$ is the covariant two time Green function
\begin{equation}
{\mathcal G}^{-1}(x,y;\chi)= [\Box + \chi(x)] \delta^2 (x-y).
\end{equation}
The gap equation  is obtained from
\begin{equation}
\frac{\partial V_{eff}}{\partial \rho} = \frac{1}{2} [\mu^2 - \chi+\lambda_0 \rho
+ \eta \rho^2]= 0
\end{equation}
and $\rho$ can be eliminated in favor of $\chi$ using
\begin{equation}
\frac{\partial V_{eff}}{\partial \chi} =0 \rightarrow \rho=\phi^2 + \frac{1}{i}
{\mathcal G}(x,x;\chi)
\end{equation}
However, $\frac{1}{i} {\mathcal G}(x,x;\chi)= G(x,x;\chi)=
\langle \{\phi(x)-\langle \phi(x) \rangle \}^2 \rangle$  where $G(x,y)$ was
defined earlier in the
variational approach. This is seen by considering
\begin{equation}
\frac{1}{i} {\cal G}(x,x;\chi)=\frac{1} {(2 \pi)^2} \int \frac {d^2
k_E}{(k_E^2 +\chi)} \equiv \frac{1}
{(2
\pi)} \int
\frac {d k}{2 \sqrt{k^2 +\chi}}
\end{equation}

 Once we rewrite $V_{eff}$ solely in terms of $\chi$ as
$V_{eff}\left[\phi^2, \chi,\rho[\chi]\right]$,
then we obtain that
\begin{equation}
\frac{\partial V_{eff}} {\partial \chi}|_{\phi} = \frac{\partial V} {\partial
\rho} \frac{\partial \rho} {\partial \chi}
\end{equation}
This yields the unrenormalized equation:
\begin{equation}
\frac{\partial V_{eff}} {\partial \chi}|_{\phi}= \frac{1}{2} [\mu^2 -
\chi+\lambda_0 \rho + \eta \rho^2][- \frac{1} {4 \pi
\chi}]
\end{equation}
using the same mass and coupling constant renormalization as in the
variational aproach we get exactly
the same result:
\begin{equation}
\frac{\partial V_{eff}} {\partial \chi}= \frac{1} {8 \pi \chi} \left[\chi -
m^2 - \lambda_r\left(\phi^2
+\frac{1}{4 \pi} {\ln} [\frac{m^2}{\chi}] \right)- \eta \left( \phi^2
+\frac{1}{4 \pi}
{\ln} [\frac{m^2}{\chi}] \right)^2  \right]
\end{equation}

Integrating this with respect to $\chi$ we obtain the renormalized effective
potential:
\begin{eqnarray}
V_{eff}[\phi,\chi] && = \frac{1}{2} m^2 \phi^2 + \frac{1}{4} \lambda_r \phi^4
+\frac{1}{6} \eta \phi^6
\nonumber \\
&& \frac{1}{8 \pi} [ \chi - m^2 + (m^2+ \lambda_r \phi^2+\eta \phi^4) \ln
\frac{m^2}{\chi} ] \nonumber
\\
&&\frac{1}{64 \pi^2} (\lambda_r  + 2 \eta \phi^2) \ln^2 \frac{m^2}{\chi}
+\frac{1}{384 \pi^3} \eta \ln^3
\frac{m^2}{\chi}.
\end{eqnarray}
We can rewrite this as:
\begin{eqnarray}
V_{eff}[\phi,\chi]=&&\frac{1}{8 \pi} [ \chi - m^2]  + \frac{1}{2} m^2 [\phi^2
+G_r[\chi]]
\nonumber \\
&& + \frac{1}{4} \lambda_r [\phi^2+G_r]^2 + \frac{1}{6} \eta[\phi^2 + G_r]^3
\end{eqnarray}
This is exactly what we got from the variational method. However now we can
systematically improve on the variational result in terms of a series in
$1/N$.

To study the critical behavior of $\phi^6$ and the stability of the large-N expansion
one needs to study the condition for the determinant of the Gaussian fluctuations to
vanish, namely:

\begin{equation}
\Sigma(0) V^{\prime \prime} [\rho] + \frac{1}{2} = 0.
\end{equation}
where
\begin{equation}
\Sigma(q=0,\chi) = \frac{1}{4 \pi \chi}.
\end{equation}
Here the potential in terms of $\rho$ is given by
\begin{equation}
V[\rho] = \frac {\mu^2}{2} \rho + \frac{\lambda_0} {4} \rho^2 + \frac{\eta}{6} \rho^3
\end{equation}
The second derivative is to be evaluated at the stationary point where
$\chi=m^2[\phi]$, $\rho = \phi^2+ G[\chi,\phi]$

We can rewrite everything in terms of renormalized parameters by having 
$\mu \rightarrow m$; $\lambda_0 \rightarrow \lambda_r$ and $\rho-> \phi^2 + G_r$
where everything is normal ordered with respect to the renormalized mass of the 
$\phi$ field $m$. 
Taking the second derivative at the stationary phase point we obtain when $\phi
\rightarrow 0$  and $\chi \rightarrow m^2$ that (since $G_r[\chi=m] =0$)
\begin{equation}
V''[\rho] = \lambda_r + \eta \rho \rightarrow \lambda_r
\end{equation}
. Thus the critical condition for the existence of a zero in the 
inverse propagator is
\begin{equation}
\frac{\lambda_r}{4 \pi m^2} +\frac{1}{2} =0.
\end{equation}
In $\phi^6$ we can have the renormalized $\lambda_r$ negative and still have a positive
definite theory as long as $\eta >0$.

\subsection{O(N) Sine-Gordon}
Here the Lagrangian divided by $N$ written in terms of the auxiliary fields
$\chi,\rho$  is given by \begin{equation}
 \frac{1}{2 N} [
\partial_\mu \phi_i ]^2-
\frac{1}{2 N } \chi {\vec \phi \cdot \vec \phi}-  V[{\rho}]
\end{equation}
where
\begin{equation} 
V[{\rho}]=-\frac{\alpha_0}{\beta^2} {\rm
cos} \beta \sqrt{\rho}
\end{equation}

Following what we did for $\phi^6$ field theory we have that the effective
potential is now:
\begin{equation}
V_{eff} =  \frac{1}{2} \chi( \phi^2 - \rho)  -\frac{\alpha_0}{\beta^2} {\rm
cos} \beta \sqrt{\rho}+ \frac{1}{2i} Tr {\rm Ln} {\cal G}^{-1}[\chi]
\end{equation}
where again  $\cal G$ is the covariant Green's function with
\begin{equation}
{\cal G} ^{-1}= [\Box + \chi] \delta^2 (x-y).
\end{equation}
The gap equation  is obtained from
\begin{equation}
\frac{\partial V_{eff}}{\partial \rho} = \frac{1}{2} [\alpha_0 \frac {\sin \beta
\sqrt{\rho}} {\beta \sqrt{\rho}} -
\chi]= 0
\end{equation}
and $\rho$ again can be eliminated in favor of $\chi$ using
\begin{equation}
\frac{\partial V_{eff}}{\partial \chi} =0 \rightarrow \rho=\phi^2 + \frac{1}{i}
{\cal G}(x,x;\chi) \end{equation}
As in the polynomial potential cases, we can render the theory finite by
just
regulating $\rho$ with respect to an arbitrary mass parameter $m_1$
\begin{equation} \rho_R=\phi^2 + \frac{1}{i} [{\cal G} (x,x;\chi)-
{\cal G(} x,x;m_1^2)] \end{equation}
so that
\begin{equation}
\rho_R = \phi^2 + \frac{1}{4 \pi} \ln[\frac{m_1^2}{\chi}]
\end{equation}
The gap equation for $\chi$ at $\phi=0$ gives the renormalized mass $m^2$:
\begin{equation}
m^2 = \alpha_0 \frac {\sin \left( \beta \sqrt{ \frac{1}{4 \pi}
\ln[\frac{m_1^2}{m^2}]} \right)}{\beta \sqrt{
\frac{1}{4
\pi}
\ln[\frac{m_1^2}{m^2}]}}
\end{equation}
so that choosing our subtraction point to be $m_1=m$ we obtain that
\begin{equation}
\alpha_0 = m^2
\end{equation}

Next we have again
\begin{equation}
\frac{\partial V} {\partial \chi}|_{\phi} = \frac{\partial V} {\partial
\rho} \frac{\partial \rho} {\partial \chi}
\end{equation}
Using our renormalization scheme $ \rho \rightarrow \rho_R$, we obtain
\begin{equation}
\frac{\partial V_{eff}} {\partial \chi}= \frac{1} {8 \pi \chi} \left[\chi -
{m^2}
\frac {\sin \left( \beta \sqrt{\phi^2+ \frac{1}{4 \pi}
\ln[\frac{m^2}{\chi}]} \right)}{\beta \sqrt{
\phi^2 +\frac{1}{4
\pi}
\ln[\frac{m^2}{\chi}]}}   \right]
\end{equation}
Integrating this with respect to $\chi$ gives the renormalized effective
potential
\begin{equation}
V[\phi,\chi] = \frac{1}{8 \pi} [\chi -m^2] -\frac{m^2}{\beta^2} \cos \left(
\beta \sqrt{\phi^2+
\frac{1}{4 \pi} \ln[\frac{m^2}{\chi}]} \right)
\end{equation}
which again displays the usual result at large-N that the renormalized
expectation of the Classical potential has the property:
\begin{equation}
\langle V[\phi^2]\rangle \rightarrow  V[\phi^2+ G_r[\chi] ]
\end{equation}

The second derivative of $V_{eff}$ with respect to $\chi$ at the minimum $\chi=
m^2$ (for the unbroken vacuum
$\phi=0$)
is
\begin{equation}
- \frac{\beta^2- 24 \pi}{192 \pi^2 \chi}.
\end{equation}
Thus the vacuum is stable as long as
\begin{equation}
\beta^2 \leq 24 \pi.
\end{equation}

Let us now show that this result is the same as demanding that the large-N expansion
be an expansion about a minimum by considering the determinant of the
fluctuations around the minimum. Critical behavior is
given by the vanishing of the determinant (Eq. \ref {eq:detzero}). We have, 
using our renormalization scheme,  \begin{equation}
V[\rho] = -\frac {\alpha_0} {\beta^2} \cos \beta \sqrt{\rho} \rightarrow V_r[\rho_r]
\end{equation}
where now
\begin{equation}
V_r[\rho_r] =-\frac {m^2} {\beta^2} \cos \beta \sqrt{\rho_r}
\end{equation}
and
$\rho_r = \phi^2 + G_r$ with the subtraction being made at the physical mass $m$.
Taking two derivatives we find that
\begin{equation}
V''[\rho] = m^2 \frac{\beta^2}{4 x^2} \left[ \cos x - \frac{\sin x}{x}     \right]
\end{equation}
where $x= \beta \sqrt \rho_r$ , $\rho_r = \frac{1}{4 \pi} \ln \frac{m^2}{\chi}$
At the stationary phase pont we need the limit of $V''$ as $x \rightarrow 0$ 
This is
\begin{equation}
V''[\rho] \rightarrow  -m^2 \frac{\beta^2}{12}
\end{equation}
The condition for the determinant of the fluctuations to be positive is 
\begin{equation}
\Sigma(0) V^{\prime \prime} [\rho] + \frac{1}{2}  >0, \label{eq:pos}
\end{equation}
where
\begin{equation}
\Sigma(q=0,\chi) = \frac{1}{4 \pi \chi}.
\end{equation}
Eq. \ref{eq:pos} leads to the condition (when $\chi \rightarrow m^2$) 
\begin{equation}
\beta^2  \leq 24 \pi.
\end{equation}
This is the same result as that we found for the stability of the vacuum 

To go beyond the lowest order approximation one first includes the Gaussian
fluctuations and
considers $\rho$ and  $\chi$ as components of a new two component field with
matrix inverse propagator $D^{-1}$.
The one particle irreducible generating functional now gets a contribution
\begin{equation}
\frac{i}{2 N} Tr Ln[D^{-1}]
\end{equation}
However this naive $1/N$ expansion is secular for time evolution problems as
shown in \cite{ref:Bogdan}.  A better approximation which avoids secularity
is a resummed next to leading order
in 1/N expansion obtained from the two-particle
irreducible formalism \cite{ref:CJT} written in terms of the fields
$\phi,\chi,\rho$ and propagators ${\cal G}$ and $D$. The  self-consistent
$1/N$ correction is found from the Graph in $\Gamma_2[\Phi, {\cal G}]$
\begin{equation} \Gamma_2= \frac{i}{4} Tr \int dx dy  {\cal G}(x,y) D(x,y)
{\cal G}(y,x) \end{equation} which will give the same Schwinger Dyson
equations
as the standard 1-PI 1/N  expansion except with FULL propagators and not
leading order ones in the self energy graphs.  This is discussed in detail
in \cite{ref:dawson} \cite{ref:berges}.
\section{Conclusions}
In this paper we have shown how to generalize the Sine-Gordon equation to the case of
O(N) symmetry.  This then allows  us to go beyond the Gaussian
approximation (at large-N) in a systematic, controlable way.  We found that at
leading order in large N, the pole in the propagator looks functionally
different from that found in the Hartree approximation for $N=1$. Thus the
combinatorics of large-N for non-polynomial potentials leads to expressions for
the evolution of the one and two point function which look qualitatively
different from what is found for polynomial potentials. We also found that the
stability of the mean-field vacuum found using a Hamiltonian approach leads to
the same condition as the stability of the large-N approximation found by
studying Gaussian fluctuations about the leading order. We briefly
discussed the fact that the naive $1/N$ expansion is secular beyond leading
order and has to be replaced by a resummed $1/N$ expansion obtainable from a
two particle irreducibility approach. 

\acknowledgments This research is supported by the DOE under contract
W-7405-ENG-36.


\begin{thebibliography}{9}

\bibitem{ref:Coleman} S. Coleman, Phys.Rev. {\bf D} 11, 2088 (1975)
\bibitem{ref:Boy} D. Boyanovsky, F. Cooper, J.J. de Vega and P.
Sodano Phys. Rev. {\bf D} 58, 025007 (1998).  hep-ph/9802277
\bibitem{div} Paolo Di Vecchia  and Moshe Moshe, Phys.Lett.B300:49-52,1993 
e-Print Archive: hep-th/9211132
\bibitem{ref:Eyal} Galit
Eyal,
Moshe Moshe, Shinsuke Nishigaki, Jean Zinn-Justin .  Nucl.Phys.B470, 369
(1996)  e-Print Archive: hep-th/9601080 
\bibitem{ref:Dirac}   P. Dirac. Proc. Camb. Phil. Soc. 26, 376 (1930)
\bibitem{ref:Hartree}  A.K. Kerman
and S.E. Koonin, Ann. Phys. 100, 332 (1976);  R. Jackiw and A. K. Kerman,  
Phys. Lett. A71, 158 (1979);
 F. Cooper, S.-Y. Pi and
P. Stancioff, Phys. Rev. D, 34,
3831 (1986); F. Cooper and S. -Y. Pi in {\em Current Trends in
Physics}
Ed. by A. Khare and T. Pradhan (World Scientific, Singapore, 1986).
 F. Cooper and E. Mottola,  Phys. Rev.  D36, 3114 (1987);
S.Y. Pi and M. Samiullah, Phys. Ref. {\bf D 36}, 3128 (1987)
 D. Boyanovsky and  H.J. de Vega,
Phys. Rev.  {\bf D47}, 2343 (1993);
 D. Boyanovsky, M. D'Attanasio, H.J. de Vega, R. Holman
Phys.Rev. D54 (1996) 1748.
\bibitem{ref:BMB} W.A.
Bardeen,M.Moshe, and M. Bander, Phys. Rev. Lett. 52,
1188 (1984)
\bibitem{ref:chang} S.J. Chang, Phys. Rev. D 13, 2778 (1976).
\bibitem{ref:Bogdan} B. Mihaila F. Cooper and J. Dawson Phys. Rev. D63 (2001) 096003.
\bibitem{ref:CJT} J. Cornwall, R. Jackiw and E. Tomboulis, Phys. Rev. D10, 2428 (1974),
Gordon Baym, 
\bibitem{ref:dawson} K. Blagoev, F. Cooper, J. Dawson and B. Mihaila Phys. Rev. D64 (2001)
125003 hep/ph 0106195.
F. Cooper, J Dawson, B. Mihaila Phys.Rev.D67:056003,2003 	
hep-ph/0209051 ibid:Phys.Rev.D67:051901,2003 ; hep-ph/0207364
\bibitem{ref:berges} J. Berges Nucl.Phys.A699:847-886,2002 
e-Print Archive: hep-ph/0105311;G. Aarts, D. Ahrensmeier, R.
Baier, J. Berges, J. Serreau Phys.Rev.D66:045008,2002 
e-Print Archive: hep-ph/0201308 .
\end{thebibliography}
\end{document}